\documentclass[prl,twocolumn,english,superscriptaddress]{revtex4-1}
\usepackage[T1]{fontenc}
\usepackage[latin9]{inputenc}
\usepackage{graphicx}
\usepackage{amsmath}  
\usepackage{xfrac}
\usepackage{txfonts}
\usepackage{float}
\usepackage[mathscr]{euscript}

\makeatletter
\usepackage{babel}
\usepackage{color}
\usepackage[colorlinks = true,
            linkcolor = blue,
            urlcolor  = blue,
            citecolor = blue,
            anchorcolor = blue]{hyperref}

\makeatother

\begin{document}





\title{Dynamical scaling as a signature of multiple phase competition in Yb$_2$Ti$_2$O$_7$}
	
\author{A. Scheie}
\email{scheieao@ornl.gov}
\affiliation{Neutron Scattering Division, Oak Ridge National Laboratory, Oak Ridge, TN 37831}

\author{O. Benton}
\affiliation{Max Planck Institute for the Physics of Complex Systems, N{\"o}thnitzer Str. 38, Dresden 01187, Germany}

\author{M. Taillefumier}
\affiliation{ETH Zurich, Swiss National Supercomputing
  Centre (CSCS), HIT G-floor Wolfgang-Pauli-Str. 27, 8093 Zurich, Switzerland}

\author{L.D.C. Jaubert}
\affiliation{CNRS, Universit\'e de Bordeaux, LOMA, UMR 5798, 33400 Talence, France}

\author{G. Sala} 
\affiliation{Spallation Neutron Source, Second Target Station, Oak Ridge National Laboratory, Oak Ridge, Tennessee 37831}

\author{N. Jalarvo} 
\affiliation{Neutron Scattering Division, Oak Ridge National Laboratory, Oak Ridge, TN 37831}
	
\author{S.M. Koohpayeh}  
\address{Institute for Quantum Matter and Department of Physics and Astronomy, Johns Hopkins University, Baltimore, MD 21218}
\address{Department of Materials Science and Engineering, The Johns Hopkins University, Baltimore, Maryland 21218, USA}

\author{N. Shannon}
\affiliation{Theory of Quantum Matter Unit, Okinawa Institute of Science and Technology Graduate University, Onna son, Okinawa 904-0495, Japan}
	
	\date{\today}


\begin{abstract}
$\rm Yb_2Ti_2O_7$ is a celebrated example of a pyrochlore magnet with highly-frustrated, anisotropic exchange interactions.
To date, attention has largely focused on its unusual, static properties, many of which can be understood as coming from the competition between different types of magnetic order.
Here we use inelastic neutron scattering with exceptionally high energy resolution to explore the dynamical properties of $\rm Yb_2Ti_2O_7$.
We find that spin correlations exhibit dynamical scaling, analogous to behaviour found near to a quantum critical point.
We show that the observed scaling collapse can be explained within a phenomenological theory of multiple--phase competition, and confirm that a scaling collapse is also seen in semi--classical simulations of a microscopic model of $\rm Yb_2Ti_2O_7$.
These results suggest a general picture for dynamics in systems with competing ground states.
\end{abstract}


\maketitle

Frustration generates competition.
When the interactions of a many body system are frustrated,
it is common to find many competing phases close in energy to the ground 
state \cite{harris97, moessner98, laeuchli19}.
Even though some particular order may emerge as the stable ground state at sufficiently low temperature, the proximity of the competing phases may still have a substantial influence on the system's properties \cite{Yan2013-preprint,andrade14, banerjee16,  Yan2017, hallas18, vojta18}.  
In such a case, we must understand the system through the lens of multiple phase competition.


\begin{figure*}[ht]  
	\centering\includegraphics[width=0.97\textwidth]{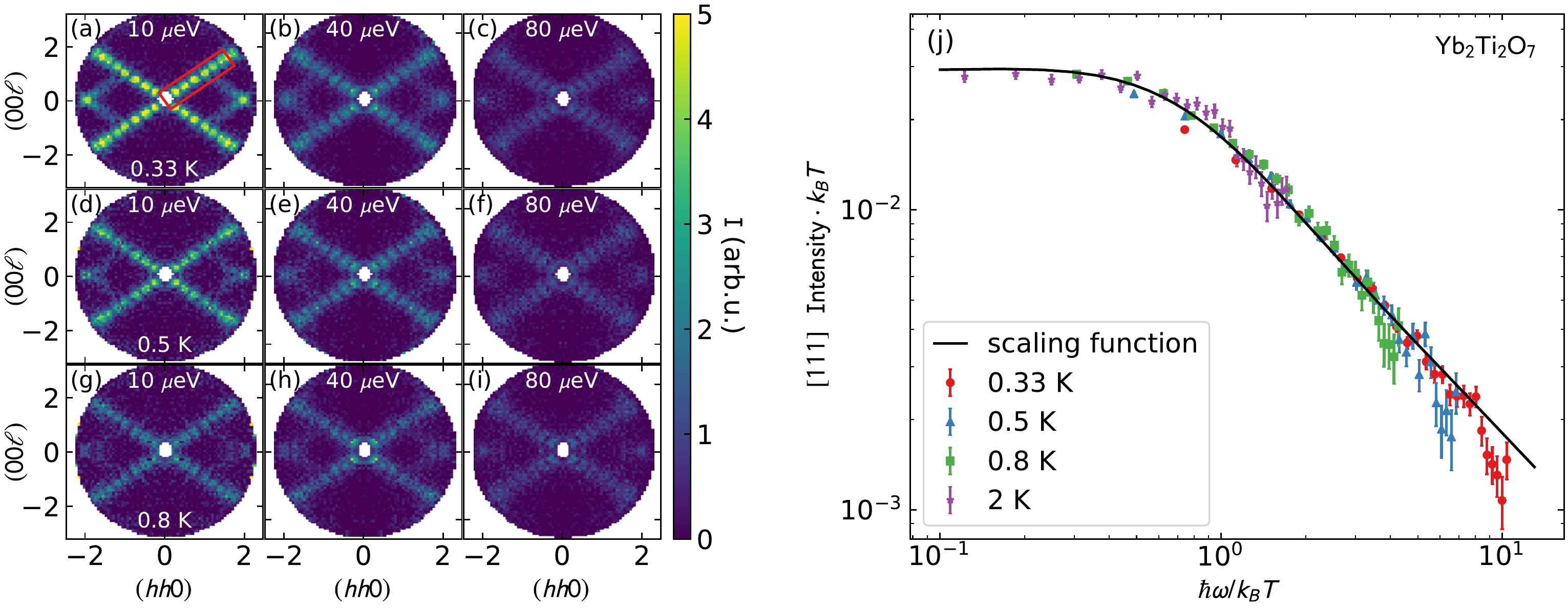}
	\caption{Low-energy neutron scattering from the short-range correlated phase of $\rm Yb_2Ti_2O_7$. Panels (a)-(i) show colorplots of neutron scattered intensity, with the horizontal rows showing three different temperatures and the vertical columns show different constant energy slices in the $hh\ell$ scattering planes. All temperatures and energies show diffuse scattering rods along $\{ 111 \}$ directions along with crosses of scattering centered at $(220)$. 
	Panel (j) shows the data integrated over the $\{ 111 \}$ scattering rods [indicated by the red box in panel (a)] scaled by the temperature. Up to 2~K, the data collapses onto itself and follows a scaling relation of type Eq.~\ref{eq:scaling_1}.
	The specific form of scaling predicted by our phenomenological theory of multiple-phase competition, Eq.~\ref{eq:scaling}, is shown with a solid line. 
	Error bars indicate one standard deviation.
	}
	\label{flo:BSSdata}
\end{figure*}

This multiple phase competition perspective has yielded especially helpful insight into rare-earth pyrochlore magnets \cite{Yan2017},  
most prominently $\rm Yb_2 Ti_2O_7$  \cite{Blote1969,HodgesCEF,Hodges2002,FM_order2003,bonville2004transitions,Ross2009,Thompson_2011,Ross_Hamiltonian,Chang,DOrtenzio_noGsOrder,FirstOrder_magnetization,ApplegateSpinIce,Robert2015,Jaubert2015,GaudetRoss_order,Yan2017,Kermarrec2017,Thompson_2017,Scheie2017,Antonio2017,rau2019magnon,Bowman2019,Saubert_2020,scheie2020multiphase,petit2020way}.
Composed of magnetic Yb$^{3+}$ ions arranged in a lattice of corner-sharing tetrahedra, the system orders ferromagnetically at $T=270$~mK \cite{SeyedPaper,Scheie2017,FirstOrderTransition}.
Its magnetic Hamiltonian lies extremely close to the boundary between canted ferromagnetic (FM) order and $\Gamma_5$ antiferromagnetic (AFM) order \cite{Ross_Hamiltonian, Robert2015, Thompson_2017, scheie2020multiphase}.
And in the broader parameter space, this phase boundary terminates in a spin liquid where it meets a $\Gamma_7$ AFM \cite{benton16-pinchline,Yan2017}. 
Various static properties of $\rm Yb_2 Ti_2O_7$, such as its low ordering temperature, the strong variation between samples and the equal-time spin correlations have been understood as arising from multiple phase competition \cite{Yan2013-preprint,Thompson_2017,scheie2020multiphase,Robert2015,Jaubert2015,Yan2017}. 
%

However not all behaviors of $\rm Yb_2 Ti_2O_7$ are well--understood, particularly those relating to dynamics.
Above the long range magnetic ordering transition $T=270$~mK and up to $T\sim 2$~K, $\rm Yb_2 Ti_2O_7$ is in a short-ranged correlated magnetic phase \cite{Blote1969}. In this temperature regime, diffuse rods of neutron scattering appear along $\{111 \}$ directions which signal structured spin correlations \cite{bonville2004transitions,Ross2009,Thompson_2011,Chang,Robert2015}. 
The presence of these rods is a signature of the proximity of AFM order, and thus falls within the picture of multiple phase competition, 
but their energy dependence remains an open issue.
Meanwhile, thermal conductivity \cite{Li_2015,Tokiwa_2016} and thermal hall conductivity \cite{hirschberger2019enhanced} reach anomalously large values in the $\rm Yb_2 Ti_2O_7$ short-range correlated phase, and terahertz spectroscopy appears to show the presence of massive magnetic quasiparticles \cite{Armitage_monopoles_2016}.
Thus the $0.27\>{\rm K} <T < 2\>{\rm K}$ magnetic state hosts exotic but poorly understood dynamics. 
This raises the question: can the intermediate temperature dynamics of $\rm Yb_2 Ti_2O_7$ be understood via multi-phase competition? 
What role, if any, does the nearby spin liquid play?
And, more generally, does multiple-phase competition imply anything universal about the dynamics of the disordered phase, in analogy with quantum criticality?

In this study, we experimentally demonstrate a
universal scaling relation for the
energy dependence of the rod-like scattering in  $\rm Yb_2 Ti_2O_7$ and connect it
with the multiple phase competition paradigm.
This is accomplished through low-energy neutron scattering measurements between 0.3~K and 2~K, using ultra-high resolution inelastic neutron spectroscopy.
The inelastic neutron scattering intensity along the $\{111\}$
directions of reciprocal space, $S_{\rm rod}(\omega)$, is described
by a scaling relation between temperature and energy:
\begin{eqnarray}
k_B T S_{\rm rod}(\omega) = f\left( \frac{\hbar \omega}{k_B T} \right).
\label{eq:scaling_1}
\end{eqnarray}
We then show how this scaling relation can be understood within a phenomenological theory of multiple phase competition, combined with Langevin dynamics.
This theory is corroborated using semi-classical Molecular Dynamics (MD) simulations of a microscopic model known to describe $\rm Yb_2Ti_2O_7$, which confirm that the scaling behavior is associated to the region of parameter space where FM and AFM orders compete. 
We thus show that multiple phase competition has universal consequences---independent of the precise Hamiltonian---in  finite-temperature dynamics.


We measured the low-energy inelastic neutron spectrum of  $\rm Yb_2 Ti_2O_7$ between 0.3~K and 2~K using the ultra-high resolution BASIS backscattering spectrometer \cite{BASISpaper} at ORNL's SNS \cite{mason2006spallation}. The sample was two single crystals grown with the traveling solvent floating zone method \cite{SeyedPaper} (the same crystals as ref. \cite{scheie2020multiphase}) co-aligned in the $(hh\ell)$ scattering plane, and mounted in a dilution refrigerator. We rotated the sample over 180$^{\circ}$ about the vertical axis, measuring the scattering up to 300 $\mu$eV (the full bandwidth of this configuration) with 3 $\mu$eV full width at half maximum energy resolution---much higher resolution than previous measurements of these features. Constant-energy slices of the data are shown in Fig. \ref{flo:BSSdata}.
We measured the $\rm Yb_2 Ti_2O_7$ spectrum at temperatures 330~mK, 500~mK, and 800~mK with 12~K background in one experiment, and then 330~mK, 2~K, 3~K with 12~K background in a second experiment with the same sample. (12~K is well into the paramagnetic phase where all spin correlations are lost, and thus makes an appropriate background for the inelastic data---see supplemental information for details \cite{SuppMat}.) 
Because of beam heating, the cryostat thermometer may differ from the actual sample temperature; accordingly, the temperature of the lowest temperature measurement was derived from a fitted Boltzmann factor for the positive and negative energy transfer scattering on the $\{ 111 \}$ feature: $T = 0.33(4)$~K.

As is clear from Fig. \ref{flo:BSSdata}, the inelastic scattering pattern in the short-range correlated phase has well-defined rods of scattering extending along $\{ 111 \}$ directions. As energy transfer $\omega$ increases, the scattering pattern grows weaker and broadens, but does not change its overall character.
Intriguingly, the same effect is observed as temperature increases: the rod scattering pattern is preserved but grows weaker and broader.
This suggests a scaling relation between temperature and energy.

To test this hypothesis, we integrated the $\{ 111 \}$ rod scattering [shown by a red box in Fig. \ref{flo:BSSdata}(a)] and plotted the intensity multiplied by temperature as a function of energy divided by temperature in Fig. \ref{flo:BSSdata}(j). We find that the data collapse onto a universal curve, and above $\hbar \omega / k_B T \approx 1$ the data follow a $(\hbar \omega / k_B T)^{-n}$ power law behavior, with a fitted exponent $n=1.03(3)$. (In the supplemental information, we show this exponent to be robust against different $Q$ integration regions \cite{SuppMat}.)
This implies a scale-invariance in the dynamics of the $\rm Yb_2 Ti_2O_7$ short-range correlated phase.

\begin{figure*}
    \centering
    \includegraphics[width=0.96\textwidth]{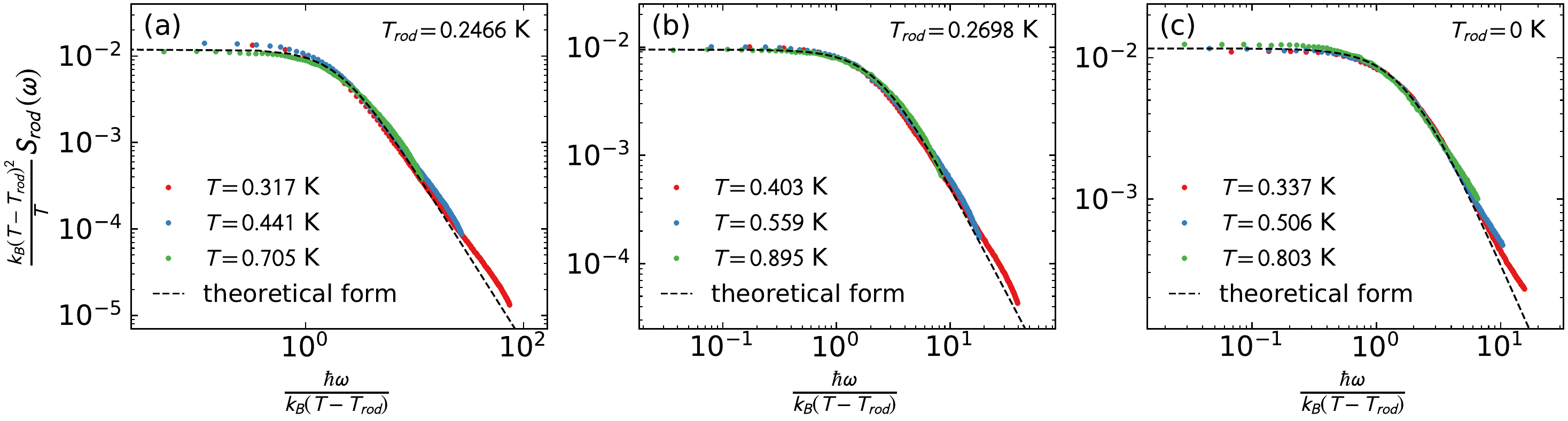}
    \caption{Dynamical scaling collapse of $S^{\rm cl}_{\rm rod}(\omega)$ calculated using
    Molecular Dynamics simulations. (a) Calculations with exchange parameters set to the values estimated for $\rm Yb_2Ti_2O_7$
    in \cite{scheie2020multiphase} (point A in Fig. \ref{fig:pd_3d}). A near, but imperfect, collapse is observed.
    (b) Calculations with a modified value of $J_1$ such that the exchange parameters lie on the FM/AFM boundary
    (point B in Fig. \ref{fig:pd_3d}). A much closer data collapse is observed compared to (a). 
    (c) Calculations at the spin liquid point 
    $J_1=J_2=J_4=0$, $J_3<0$
    (point C in Fig. \ref{fig:pd_3d}). The collapse is observed with a vanishing value of the rod criticality temperature $T_{\rm rod}$.}
    \label{fig:simulations}
\end{figure*}


To understand this, we construct a phenomenological theory which takes into account the competition between FM and AFM phases.
Writing a Ginzburg-Landau theory with dissipative dynamics \cite{Hohenberg77} in terms of competing order parameters of ferromagnetic and antiferromagnetic phases, and assuming low-energy modes along $\{111\}$ which collapse to zero energy at some temperature $T_{\rm rod}$, we find an equation (derived in the Supplemental Information \cite{SuppMat}) for the inelastic structure factor of a $[111]$ rod $S_{\rm rod}$ 
%
\begin{eqnarray}
&&S_{\rm rod} (\omega)
= \int_{q_1}^{q_2}
S(q_{111}, \omega)
\nonumber \\
&&
=
2  (n_{BE}(\omega) + 1)\frac{1}{k_B^2 (T-T_{\rm rod})^2}
\frac{B \hbar \omega}{R^2 + \frac{\hbar^2\omega^2}{k_B^2 (T-T_{\rm rod})^2}}.
\label{eq:FullScalingRelation}
\end{eqnarray}
Here $B$ and $R$ are non-universal dimensionless constants, and $n_{BE}(\omega)$ is the Bose-Einstein distribution. 

Fitting Eq. \ref{eq:FullScalingRelation} to the $\rm Yb_2Ti_2O_7$ experimental data, we find good agreement with $T_{\rm rod} = -0.05(5)$. This is zero to within uncertainty. Setting $T_{\rm rod}=0$ explicitly we obtain the scaling relation (\ref{eq:scaling_1}), with the scaling function:
\begin{eqnarray}
f(x)
=2 
\left(\frac{1}{\exp(x)-1}  + 1 \right)
\frac{B x}{R^2 +
x^2
}
\label{eq:scaling}
\end{eqnarray}
which depends only on the ratio
$x=\frac{\hbar \omega}{k_B T}$.
This form for $f(\frac{\hbar \omega}{k_B T})$ beautifully matches the experimental data as shown in Fig. \ref{flo:BSSdata}, with fitted constants $B=0.0181(3)$ and $R=0.80(3)$.

The crucial ingredients in the phenomenological theory behind Eq. \ref{eq:scaling} are (i) dissipative dynamics; (ii) close competition between two phases, here ferromagnetic and antiferromagnetic;
(iii) flat, low energy modes, along the $\{111\}$ directions;
(iv) a collapse of these modes to zero energy at some temperature $T_{\rm rod}$;
(v) $T_{\rm rod}\approx0$.

Of these, (i) is a natural assumption for a paramagnetic phase in a strongly interacting system, (ii) has been inferred previously from the static behavior of Yb$_2$Ti$_2$O$_7$ \cite{Jaubert2015, Robert2015, scheie2020multiphase} and (iii) is known to follow from (ii) \cite{Yan2017}.
Explaining the data then requires one novel assumption [(iv)] and an empirical determination that $T_{\rm rod}\approx0$ for $\rm Yb_2Ti_2O_7$ \cite{SuppMat}.

To validate the idea of a
temperature-dependent, collapsing, energy scale for the $\{111\}$ rods in a microscopic model appropriate to Yb$_2$Ti$_2$O$_7$, we turn to molecular dynamics (MD) simulations.
We simulate a nearest-neighbor anisotropic exchange Hamiltonian:
\begin{eqnarray}
H_{\rm ex}=
\sum_{\langle ij \rangle} \sum_{\alpha \beta} J_{ij}^{\alpha \beta} S_i^{\alpha} S_j^{\beta}.
\label{eq:H_ex}
\end{eqnarray}
The form of the exchange matrices 
$J_{ij}^{\alpha \beta}$ is fixed by symmetry \cite{Curnoe_2007, Ross_Hamiltonian, Yan2017}
and there are four independent parameters $\{J_k\}=\{J_1, J_2, J_3, J_4\}$.
Several different estimates of these parameters are available for Yb$_2$Ti$_2$O$_7$ 
\cite{Ross_Hamiltonian, Robert2015, Thompson_2017, scheie2020multiphase}, generally placing 
Yb$_2$Ti$_2$O$_7$ close to a phase boundary between ferromagnetic and antiferromagnetic order \cite{Yan2017}.

The dynamics of the model (Eq. \ref{eq:H_ex})
are simulated following the method in (e.g.) \cite{conlon09, taillefumier14, taillefumier17}. First, an ensemble of configurations is generated at temperature $T$ using a classical Monte Carlo simulation, treating the spins as vectors of fixed length $|{\bf S}_i|=1/2$. 
We then time-evolve the configurations 
using the Heisenberg equation of motion
\begin{eqnarray}
\hbar \partial_t {\bf S}_i (t) =
{\bf S}_i (t) \times {\bf h}^{\rm eff}_i (t)
\label{eq:LandauLifshitz}
\end{eqnarray}
where ${\bf h}^{\rm eff}_i (t)$ is the effective exchange field produced by the  spins surrounding $i$.
The dynamical structure factor is 
then calculated by Fourier transforming the correlation functions in both time and space and averaging over the ensemble. 
We do not include an explicit dissipation term
in Eq. (\ref{eq:LandauLifshitz}) but the resultant 
dynamics can nevertheless be dissipative, due to
the strong interactions between modes, arising from non-linearity.

Since the simulations sample from a classical ensemble of states, the comparison of the
phenomenological theory with 
the simulation results requires using the classical fluctuation-dissipation relationship $S(\omega)=\frac{2 k_B T}{\omega} {\rm Im}[\chi(\omega)]$, as opposed to the quantum
relationship $
S(\omega)= 2  (n_{BE} (\omega) + 1 ) {\rm Im}[\chi(\omega)]
$ used to derive Eq. (\ref{eq:FullScalingRelation})
\cite{SuppMat}.
This leads to the following modified scaling law:
\begin{eqnarray}
\frac{k_B (T-T_{\rm rod})^2}{T}
S^{\rm cl}_{\rm rod}(\omega)
= \frac{A}{W^2 + \left(\frac{\hbar \omega}{k_B(T-T_{\rm rod})} \right)^2}
\label{eq:classical_scaling}
\end{eqnarray}
$S^{\rm cl}_{\rm rod}(\omega)$ is the semi-classical structure factor integrated along a $\{111\}$ rod and the right hand side of Eq. (\ref{eq:classical_scaling}) is only a function of the ratio
$ \frac{\hbar \omega}{k_B(T-T_{\rm rod})}$.
$A$ and $W$ are non-universal constants.

In Fig. \ref{fig:simulations} we show the scaling collapse of the simulated $S^{\rm cl}_{\rm rod} (\omega)$ for three different sets of exchange parameters $\{ J_k \}$.
For each parameter set, $T_{\rm rod}$ is treated as an adjustable parameter to optimize the data collapse.

In Fig. \ref{fig:simulations}(a) we show the simulation data for the exchange parameters estimated for Yb$_2$Ti$_2$O$_7$ in \cite{scheie2020multiphase}.
This parameter set lies close to the FM/AFM boundary, but not exactly on it.
Accordingly, the collapse of the simulation data is close, but imperfect.
Adjusting the value of $J_1$, such that the parameters lie exactly on the $T=0$ FM/AFM phase boundary, greatly improves quality of the data collapse as shown in Fig. \ref{fig:simulations}(b).
Moving away from the phase boundary the collapse becomes worse (see Supplemental Materials \cite{SuppMat}).
This confirms the connection between the observed dynamical scaling and the proximity of the FM/AFM phase boundary.

The MD data collapses in Fig. \ref{fig:simulations}(a)
and (b) both use finite values of
$T_{\rm rod}$.
In both cases $T_{\rm rod} < T_{\rm order}$ where $T_{\rm order}$ is the temperature of a magnetic ordering transition.
Similarly, in experiment $T_{\rm rod}=0 < T_{\rm order} =0.27 \> {\rm K}$.
The point where the rods become critical is thus hidden beneath a thermodynamic phase transition and never reached in the simulations, although its effects are seen in the correlated paramagnetic phase.

\begin{figure}
    \centering
    \includegraphics[width=0.47\textwidth]{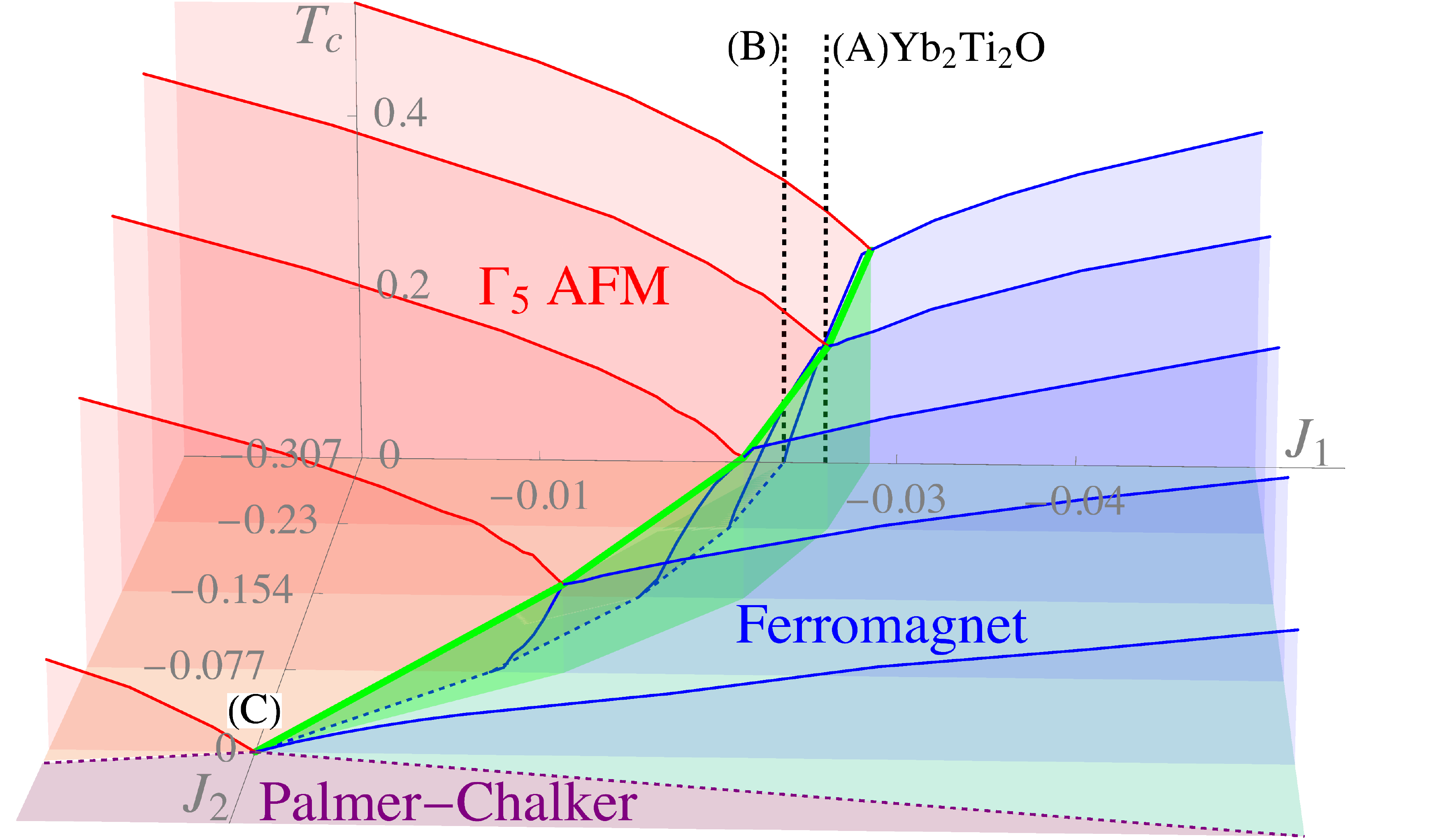}
    \caption{Finite temperature phase diagram of the pyrochlore  $\{J_1, J_2, J_3, J_4 \}$ exchange model  \cite{Curnoe_2007, Ross_Hamiltonian, SuppMat}, determined from classical Monte Carlo simulations. The horizontal axis is $J_1$, the vertical axis is temperature, and the out-of-the-page axis is $J_2$, with $J_3=-0.322$~meV and $J_4=-0.091 J_2$. The solid lines show $T_{\rm order}$ as a function of $J_1$, for a series of values of $J_2$. 
    Point A shows the $\rm Yb_2Ti_2O_7$ exchange parameters \cite{scheie2020multiphase}, Point B has the same values of $J_{2,3,4}$ as A, but $J_1$ is adjusted so as to lie exactly on the phase boundary, and Point C is a point at which FM and $\Gamma_5$ orders meet another form of antiferromagnetic order and a spin liquid emerges
    \cite{benton16-pinchline}.
    The green line shows the finite temperature boundary between the ferromagnet  (FM) and antiferromagnetic $\Gamma_5$ (AFM) states, which goes to zero at the classical pinch line spin liquid point. Thus in the finite temperature regime, $\rm Yb_2Ti_2O_7$ is continuously connected to a zero temperature spin liquid phase.
    }
    \label{fig:pd_3d}
\end{figure}

\begin{figure}[t]
    \centering
    \includegraphics[width=0.44\textwidth]{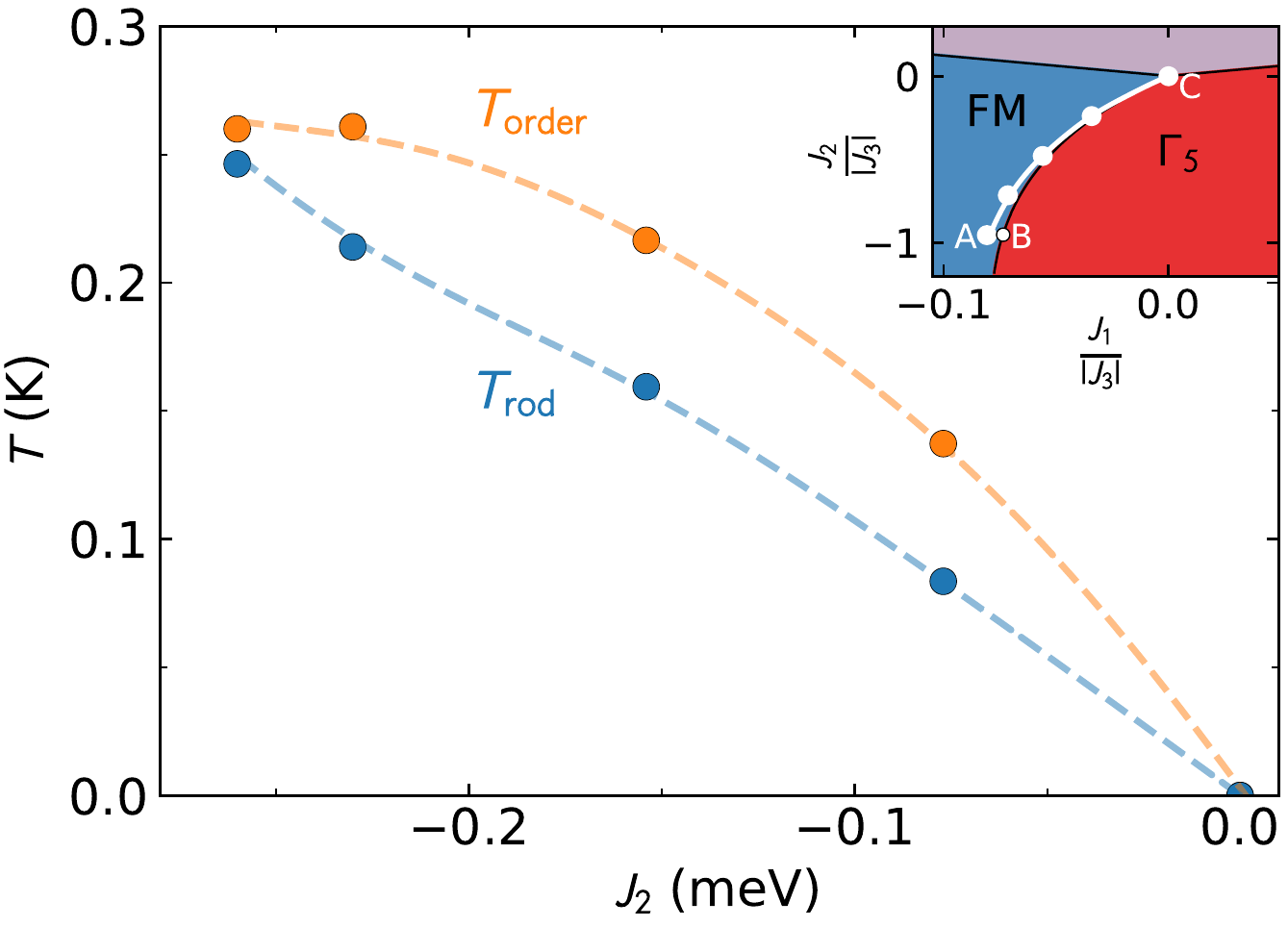}
    \caption{
    Variation of thermodynamic transition temperature
    $T_{\rm order}$, and dynamic criticality temperature $T_{\rm rod}$,  
    found in simulation.
    Results are shown for a path in parameter space which connects 
    $\rm Yb_2Ti_2O_7$ (A) to the spin-liquid point $J_2=0$ (C), 
    shown by a white line in the inset.
    Both $T_{\rm order}$ and $T_{\rm rod}$ tend to zero approaching 
    the spin liquid.
    $T_{\rm rod} < T_{\rm ordering}$ for all parameters, meaning that the 
    approach to criticality on the rods is cut-off by the ordering transition 
    as temperature is lowered.
    The effects of this hidden critical point are nevertheless seen 
    in the paramagnetic phase.
    }
    \label{fig:Trod}
\end{figure}

A striking aspect of the experimental results
is the vanishing value of $T_{\rm rod} \approx 0$,
whereas the simulations for parameters close to 
$\rm Yb_2Ti_2O_7$ find a finite value of $T_{\rm rod}$.
The vanishing of $T_{\rm rod}$ is suggestive of the influence of a spin liquid, and indeed there is such a spin liquid on the phase diagram where three ordered phases meet  and magnetic order is completely suppressed \cite{benton16-pinchline,Yan2017}. 
In Fig.~\ref{fig:pd_3d} we show how the transition temperatures of FM and AFM 
phases found in simulation collapse approaching this point, marked C. 
The temperature scale $T_{\rm rod}$ also tends to zero 
approaching the spin liquid, as shown in Fig. \ref{fig:Trod}. 
%
%
The vanishing value of $T_{\rm rod}$ in experiment is therefore suggested to stem from the influence of a nearby spin liquid, whose regime of influence is widened by quantum fluctuations.

This hypothesis, that the $\rm Yb_2Ti_2O_7$ finite temperature phase is driven by a proximate spin liquid, is reasonable given (i) the finite-temperature regime is continuously connected to the zero-temperature spin liquid, with a smooth decrease of $T_{\rm order}$ connecting the two [Fig. \ref{fig:pd_3d}], (ii) the observed experimental scaling collapse in $S_{\rm rod}(\omega)$ with $T_{\rm rod}=0$ [Fig. \ref{flo:BSSdata}.(j)] is a feature of the pinch-line spin liquid point [Fig. \ref{fig:simulations}.(c)], (iii) 
spin-wave calculations suggest that
quantum fluctuations expand the pinch-line spin liquid to a finite region in parameter space extending especially along the FM/AFM phase boundary \cite{Yan2017}. This may explain the anomalous transport behavior in the $\rm Yb_2Ti_2O_7$ finite-temperature phase \cite{Li_2015,Tokiwa_2016,hirschberger2019enhanced,Armitage_monopoles_2016}.

In summary, we have experimentally demonstrated a dynamical scaling relation in the structure factor for inelastic neutron scattering in $\rm Yb_2Ti_2O_7$.
We have shown how this scaling can be understood using a phenomenological theory based on multiple phase competition,
and demonstrated that equivalent scaling can be found in simulations of a microscopic model of $\rm Yb_2Ti_2O_7$.
These results show how multiple phase competition can have universal consequences beyond the ground state,
manifesting in the spin dynamics of a correlated
paramagnetic phase.
%



The short-range correlated phase of Yb$_2$Ti$_2$O$_7$ is thus best understood in 
terms of an underlying competition between 
ferromagnetism and antiferromagnetism and the influence of this competition extends not just to static but also dynamic properties.
The description of the dynamics in terms of a Langevin equation suggests an absence of 
long-lived propagating quasiparticles in the paramagnetic regime.
Future work will be needed to address whether this theory can explain other mysterious intermediate-temperature behaviors of $\rm Yb_2Ti_2O_7$, such as  transport.

Since extended low energy modes are quite a common
feature of frustrated magnets in general
it seems likely that a similar framework
may apply to several materials.
In particular, given that a finite-temperature correlated phase is a feature of many Yb$^{3+}$ pyrochlores \cite{hallas2018experimental}, the phenomenology seen here may prove generic to the entire class, particularly 
$\rm Yb_2Ge_2O_7$ which also lies close to a phase boundary \cite{dun15, sarkis20}. 
Moreover, since extended degenerate modes emerge on several
phase boundaries of the pyrochlore 
anisotropic exchange model (Eq. \ref{eq:H_ex}) \cite{Yan2017},  it would be interesting to search for dynamical scaling behavior in other pyrochlore oxides such as $\rm Er_2Sn_2O_7$  \cite{guitteny13, petit17-er2sn207, yahne21}.

Taking a wider perspective, our experimental results
and their interpretation via Eq. (\ref{eq:scaling})
imply an emergent relaxation time 
$\tau_{\sf rod} = \frac{1}{R} \frac{\hbar}{k_B T}$ with $R\approx 0.8$
\cite{SuppMat}. 
This is close  to the ``Planckian'' dissipation time
$\tau_{\rm Planck}=\frac{\hbar}{k_B T}$
which has been discussed as a possible fundamental bound on dissipative timescales in strongly coupled systems \cite{zaanen04, lucas19bound, bruin13resistivity, legros19resistivity}.
Experimental efforts in this area have focussed principally on charge scattering in metals, but if there is a universal principle at play it should presumably show up in other contexts too, including the spin dynamics of correlated insulators.
Whether there is any link
between these concepts and the physics uncovered here in $\rm Yb_2Ti_2O_7$
is a direction worth exploring.

\subsection*{Acknowledgments}
 This research used resources at the Spallation Neutron Source, a DOE Office of Science User Facility operated by the Oak Ridge National Laboratory. 
 A. S. acknowledges helpful discussions with D.A. Tennant.
O. B. acknowledges useful discussions 
with Benedikt Placke and Shu Zhang.
L.D.C.J. acknowledges financial support from CNRS (PICS No. 228338) and from the ``Agence Nationale de la Recherche'' under Grant No. ANR-18-CE30-0011-01. 
Single crystal development was supported as part of the Institute for Quantum Matter, an Energy Frontier Research Center, funded by the U.S. Department of Energy, Office of Science, Office of Basic Energy Sciences, under Award DE-SC0019331.
This work was supported by the Theory of Quantum Matter Unit of the 
Okinawa Institute of Science and Technology Graduate University (OIST).

%

\pagebreak

\section*{Supplemental Information for Dynamical scaling as a signature of multiple phase competition in Yb$_2$Ti$_2$O$_7$}

\renewcommand*{\citenumfont}[1]{S#1}
\renewcommand*{\bibnumfmt}[1]{[S#1]}
\renewcommand{\thefigure}{S\arabic{figure}}
\renewcommand{\thetable}{S\arabic{table}}
\renewcommand{\theequation}{S.\arabic{equation}}
\renewcommand{\thepage}{S\arabic{page}}  
\setcounter{figure}{0}
\setcounter{page}{1}
\setcounter{equation}{0}

	\section{Experiments}
	
\begin{figure*}
		\centering\includegraphics[width=0.99\textwidth]{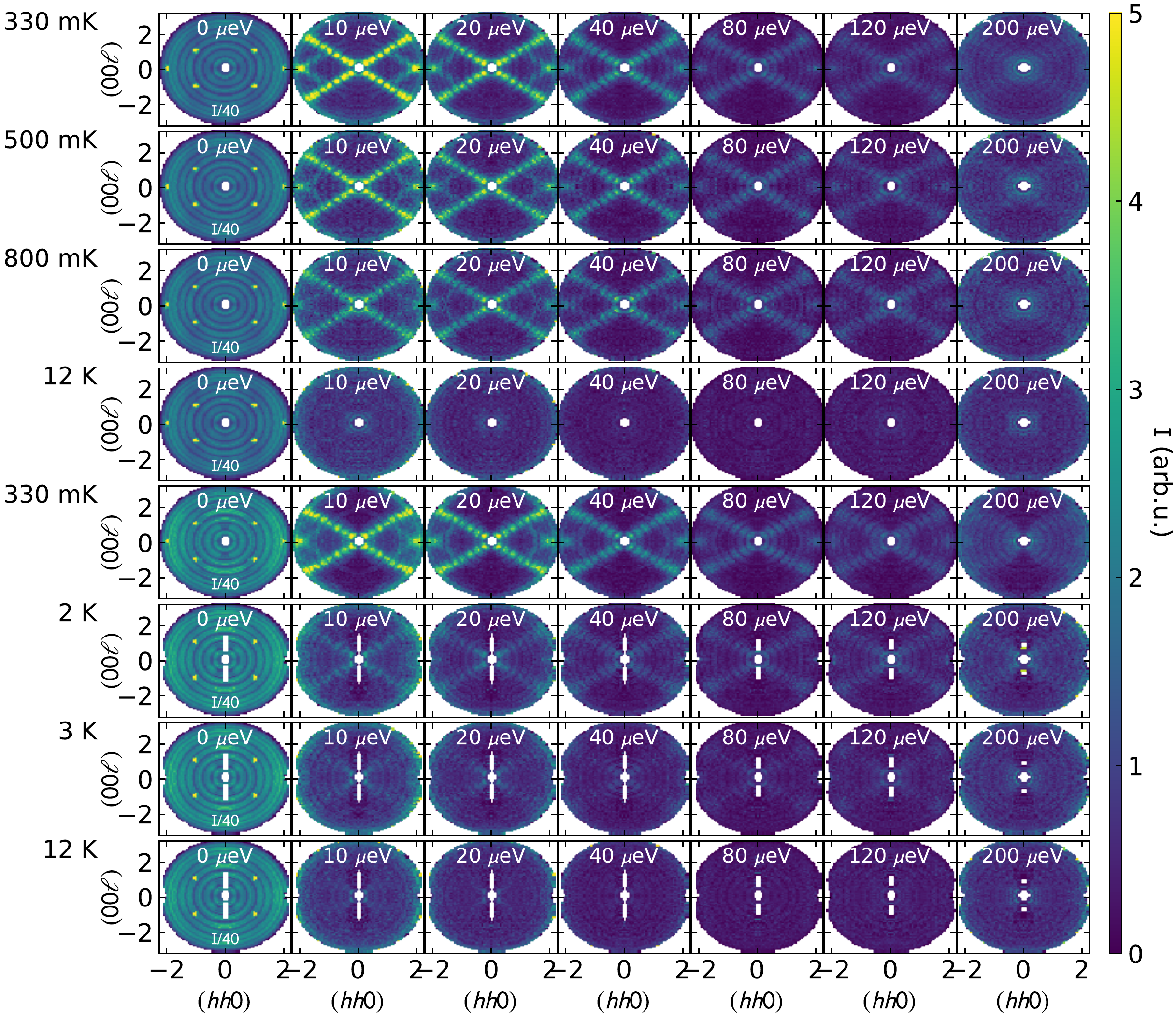}
		
		\caption{Low-energy neutron scattering from the short-range correlated phase of $\rm Yb_2Ti_2O_7$. The horizontal rows show three different temperatures, and the vertical columns show different constant energy slices in the $hh\ell$ scattering planes. The top four rows show the scattering from the first BASIS experiment, and the bottom four rows show scattering from the second BASIS experiment. All temperatures and energies show diffuse scattering rods along $\{ 111 \}$ directions along with crosses of scattering centered at $(220)$.
		}
		\label{flo:BSSdata3}
	\end{figure*}

	We measured the Yb$_2$Ti$_2$O$_7$ inelastic scattering using the Si (111) reflection on the Basis spectrometer in 60 Hz operation mode, giving a wavelength $\lambda = 6.1 \> \AA$, an energy resolution of 3.6~$\mu$eV, an energy bandwidth from -50 to $300 \> \mu$eV  \cite{BASISpaper}. The raw data from these measurements are shown in Fig. \ref{flo:BSSdata3}. As noted in the main text, these data were taken over two separate BASIS experiments using the same sample, but with slightly different slit configurations. Thus we normalized the intensity scale by the lowest temperature 10~$\mu$eV rod scattering, where the rods are strongest and clearest.
	
\begin{figure*}
		\centering\includegraphics[width=0.95\textwidth]{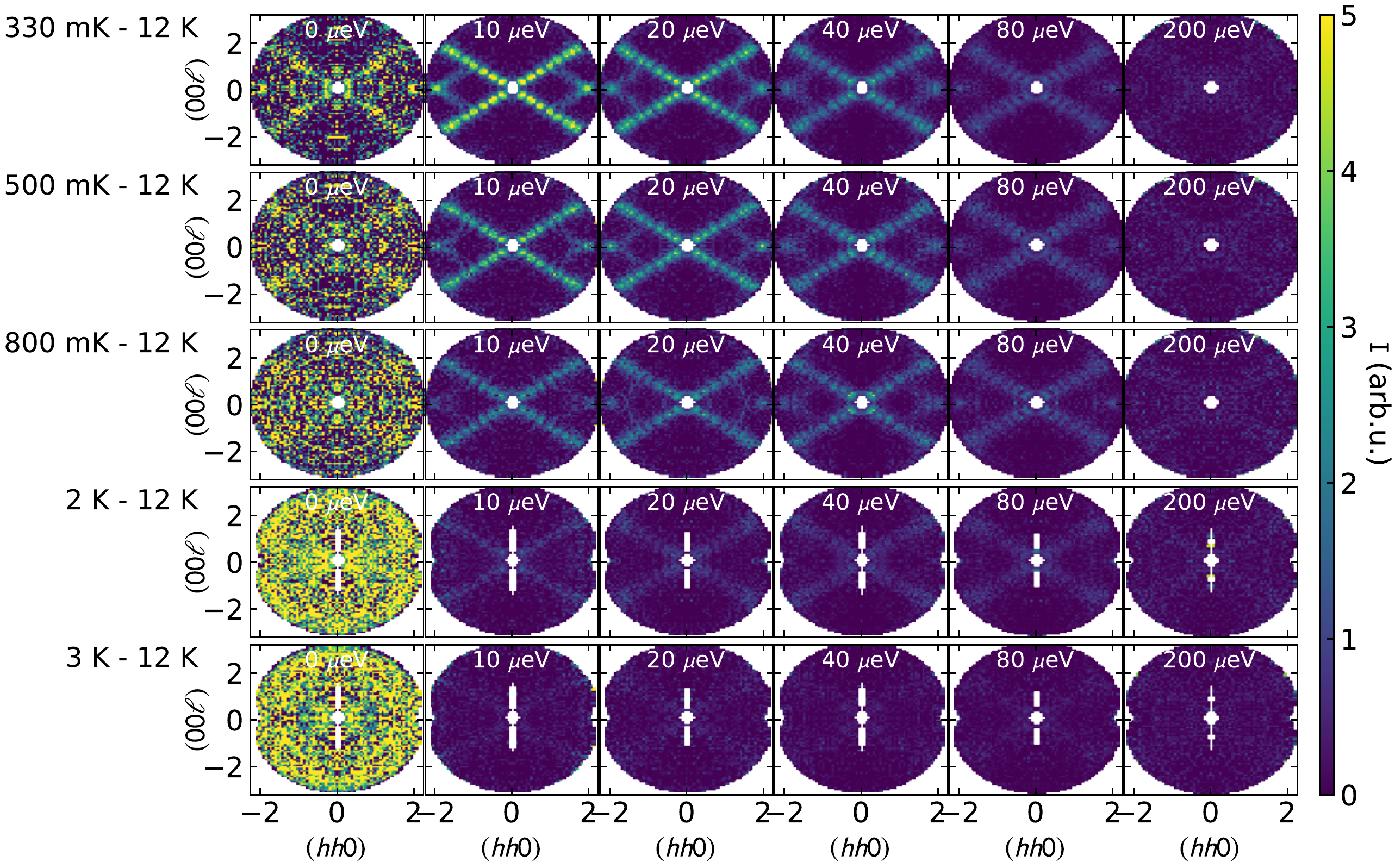}
		
		\caption{The same data as Fig. \ref{flo:BSSdata3}, but with the 12~K data subtracted as a background. The elastic signal is very noisy due to the large background, but the inelastic signal is clear at low energies. The diffuse scattering pattern broadens and grows weaker as $T$ and $\hbar \omega$ increase, in accord with the scaling relation we observe.
			In the 3~K data, the rods of intensity are barely visible, and we found that the 3~K data does not follow the critical scaling relation seen at 2~K and below.
		}
		\label{flo:BSSdata2}
	\end{figure*}

	To remove experimental artifacts from the data, we treated the 12~K scattering---well above the short-range correlated phase---as background and subtracted it from the lower temperature data. At these low energies, this eliminates artifacts isolates the magnetic scattering very well, as shown in Fig. \ref{flo:BSSdata2}. The $Q$ resolution of BASIS is not very good, which leads to a choppiness in the data in the $Q$-dependence of the rods. However, the BASIS energy resolution is excellent. Provided a large enough $Q$ region is integrated over, the magnetic scattering energy dependence is revealed in exquisite detail.
	
	\subsection{Fitting the timescale}
	
\begin{figure}
		\centering\includegraphics[width=0.49\textwidth]{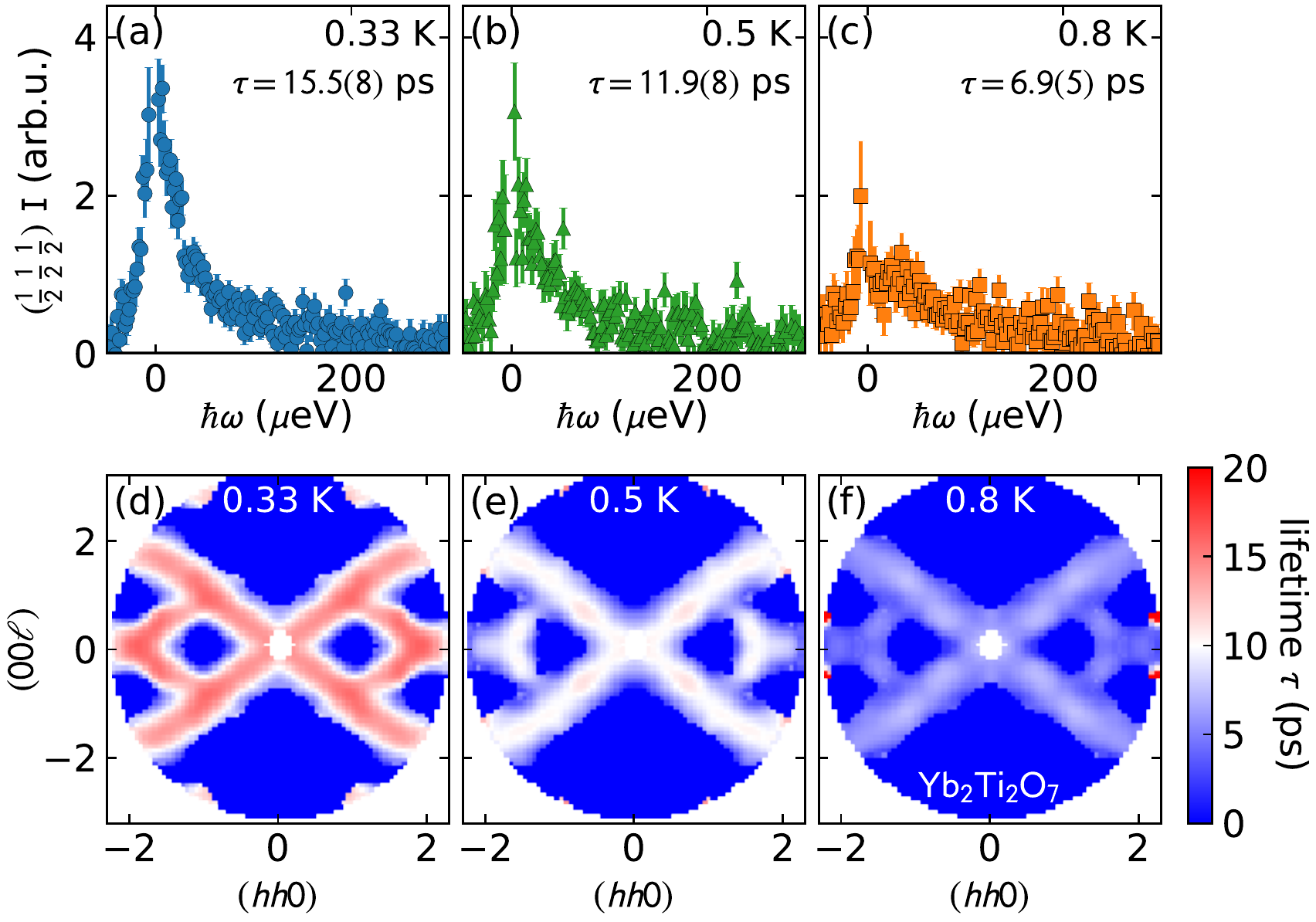}
		
		\caption{Fitted excitation lifetimes of $\rm Yb_2Ti_2O_7$ at 0.33~K, 0.5~K, and 0.8~K. The top row (a)-(c) shows the fitted function (solid black line) at a single $Q$ pixel: $Q=(\frac{1}{2},\frac{1}{2},\frac{1}{2})$. The bottom row (d)-(f) shows the fitted lifetime of every pixel above an intensity threshold. Note that the spin dynamics slow down appreciably as temperature decreases, and that the various features in the $(hh\ell)$ plane all have roughly the same lifetime regardless of intensity or wavevector.
		}
		\label{flo:FitLifetime}
	\end{figure}

	The Yb$_2$Ti$_2$O$_7$ short-range correlated phase scattering pattern has no noticeable dispersion, but does have a monotonic decrease in intensity as energy increases.
	On the simplest level, the fluctuation timescale can be extracted from the energy-dependence by fitting it to a Lorentzian function.
	We do this, pixel-by-pixel, for the 0.33~K, 0.5~K, and 0.8~K scattering in Fig. \ref{flo:FitLifetime}. In this case we use a Lorentzian function (the Fourier transform of exponential decay) weighted by the Boltzmann factor, such that the temperature can also be extracted from the imbalance between negative and positive energy transfer scattering. (Incidentally, the data in Fig. \ref{flo:FitLifetime}(a) is what was used to define the lowest temperature $T = 0.33(4)$~K.) After smoothing the data (to cover over the gaps in intensity along the \{111\} rods), the spin fluctuation timescale can be straightforwardly fitted. As expected, the fluctuation timescale decreases as temperature increases.
	
	Intriguingly, at the lowest temperatures the timescale does not appear to be correlated with the strength of the scattering feature. As shown in Fig. \ref{flo:FitLifetime}(d), at 0.33~K the \{111\} scattering rods have the same timescale ($\sim 2.5$~ps) as the weaker crosses at $(220)$. However, this $Q$-independent timescale does not appear to hold as temperature increases: by 0.8~K the $(220)$ correlations appear to fluctuate faster than the \{111\} rods, suggesting that the \{111\} rods are associated with the more persistent spin correlations at higher temperatures.
	This comports with the theoretical result that the $(220)$ feature does not have the same critical scaling as the  \{111\} rods. (Unfortunately, the statistics of the experimental $(220)$ scattering are too weak to reliably test for critical scaling directly.)
	
	\subsection{Scaling relation}
	
\begin{figure}
		\centering\includegraphics[width=0.47\textwidth]{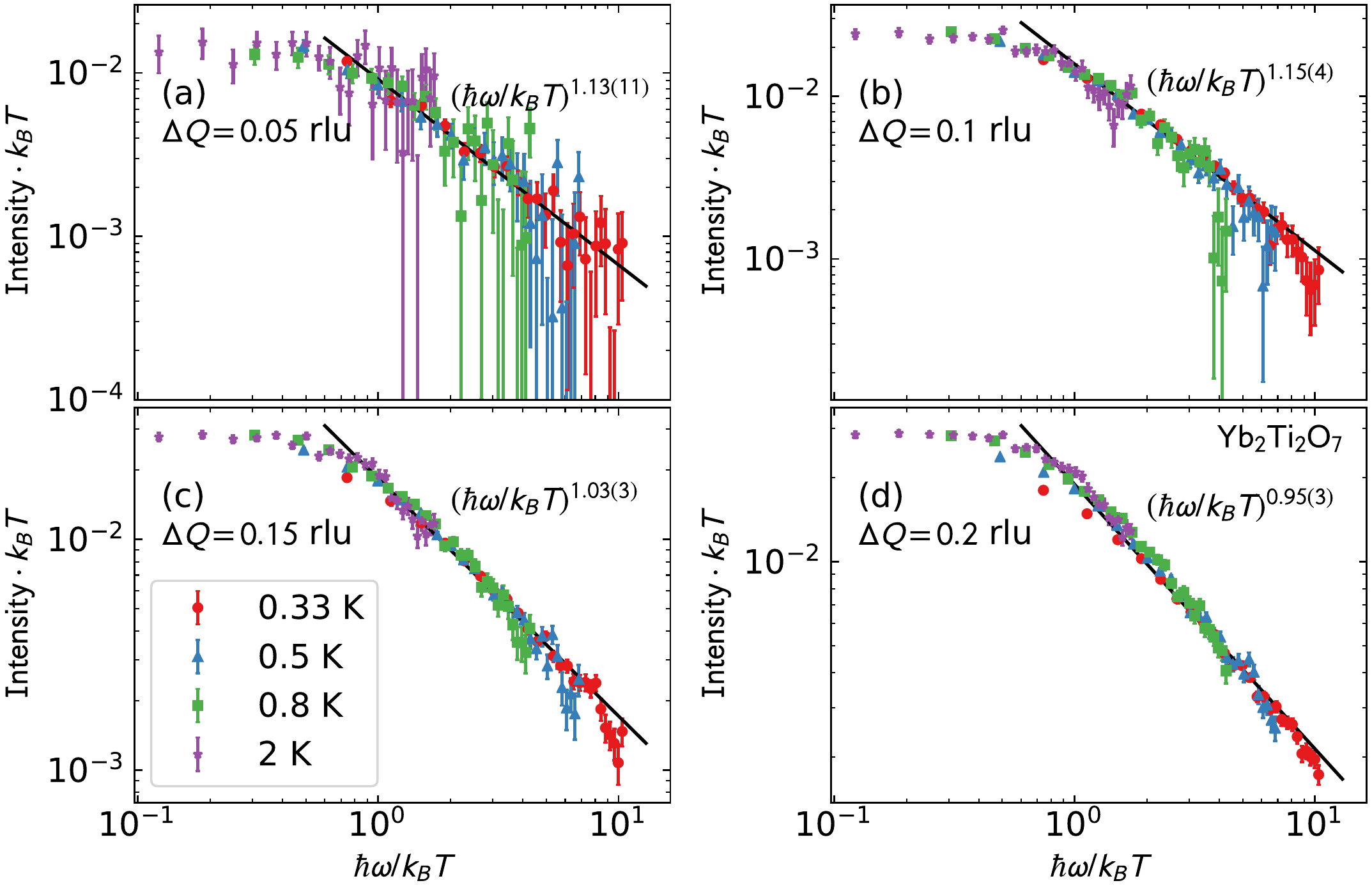}
		
		\caption{$\rm Yb_2Ti_2O_7$ scaling relation of the  \{111\} rods as a function of the width of the integrated window. Qualitatively, the results do not change as a function of integration window. rlu is reciprocal lattice units.
		}
		\label{flo:ScalingWidths}
	\end{figure}

	In the main text we show that the \{111\} rods follow a $(\hbar \omega / k_B T)^{-1}$ scaling relation. Here we show this result is robust to different integration windows. In Fig. \ref{flo:ScalingWidths}, we show the data collapse of the $\rm Yb_2Ti_2O_7$ \{111\} scattering with different integration widths in $Q$ perpendicular to the rods. As the window narrows, the data becomes considerably noisier, but it continues to follow a power law $n\approx -1$. As the window widens, the low temperature low energy scattering begins to deviate from the scaling collapse because the bin width is wider than the actual rod, making the integrated intensity smaller than it should be. Nevertheless, the overall trend still shows a $n\approx -1$ scaling relation. As a compromise between these two effects (noisy data and unreliable low-energy points) we chose to display the $\Delta Q = 0.15$~rlu (reciprocal lattice units) in the main text. 
	
\begin{figure}
		\centering\includegraphics[width=0.44\textwidth]{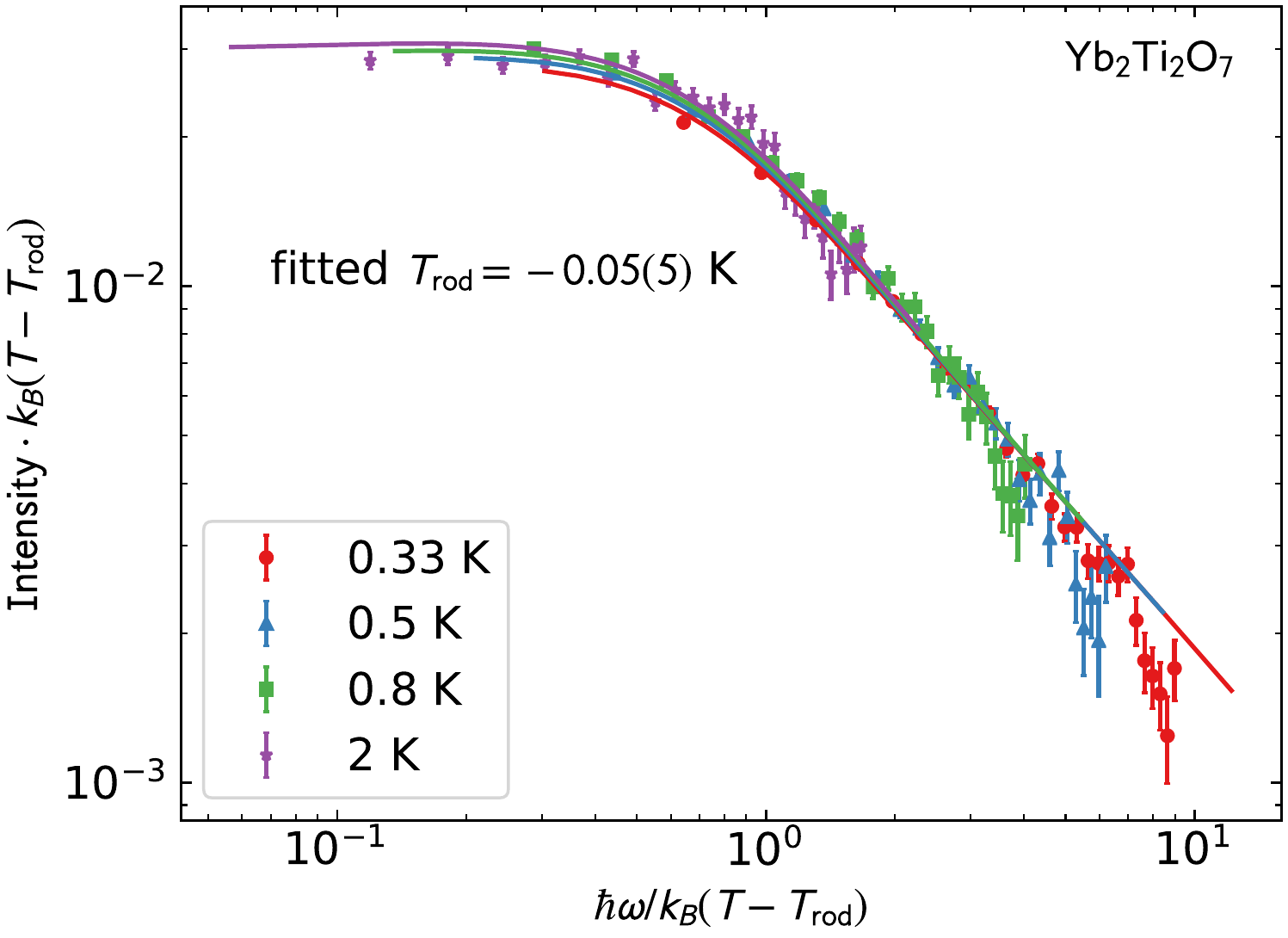}
		
		\caption{Scaling relation of the $\rm Yb_2Ti_2O_7$  \{111\} rods where $T_{\rm rod}$ (main text Eq. 2) is allowed to vary as a fitted parameter. The results show a slight deviation at low $\omega/T$, but the fitted $T_{\rm rod}$ value is negative and zero to within uncertainty. Error bars indicate one standard deviation.
		}
		\label{flo:ScalingRelationFitTrod}
	\end{figure}

	Eq. 2 of the main text gives a general equation for $\{ 111 \}$ scaling as a function of $T-T_{\rm rod}$, where the $\{ 111 \}$ rods are gapped at $T=0$ but collapse to zero at some finite temperature  $T_{\rm rod}$ [see Eq. (\ref{eq:mode_collapse})]. 
	We treated $T_{\rm rod}$ as a fitted parameter and fit the $\rm Yb_2Ti_2O_7$ $\{ 111 \}$ scattering to Eq. 2, and the results are shown in Fig. \ref{flo:ScalingRelationFitTrod}. The experimental $T_{\rm rod} = -0.05(5)$ is negative but overlaps with 0 to within one standard deviation uncertainty (it is not even clear what a negative $T_{\rm rod}$ would mean in the phenomenological theory anyway). Therefore, in our analysis of the $\rm Yb_2Ti_2O_7$ critical scaling we assume $T_{\rm rod}=0$, which gives a very good account of the data.
	
	\section{Phenomenological Theory}
	
	Here we derive scaling relation Eq. 2 in the main text.
	
	First, we write down a Ginzburg-Landau free energy
	in terms of the order parameters ${\bf m}_{\sf T_1}$
	and ${\bf m}_{\sf E}$ \cite{Yan2017}. ${\bf m}_{\sf T_1}$ is the
	three-component order parameter of the ferromagnetic
	phase, ${\bf m}_{\sf E}$ is the two component order
	parameter of the antiferromagnet \cite{Yan2017}.
	\begin{eqnarray}
	&&F_0 = \frac{1}{2} \int d^3 {\bf r} 
	\bigg[ c_{T_1} {\bf m}_{\sf T_1}^2 + c_{\sf E} {\bf m}_{\sf E}^2
	+ f_1 \left( \nabla \cdot {\bf m}_{\sf T_1} \right)^2
	+ f_2 \left( \nabla \times {\bf m}_{\sf T_1} \right)^2 \nonumber \\
	&& \qquad + f_3
	\bigg|\begin{pmatrix}
	\partial_y m_{\sf T_1}^z + \partial_z m_{\sf T_1}^y \\
	\partial_z m_{\sf T_1}^x + \partial_x m_{\sf T_1}^z \\
	\partial_x m_{\sf T_1}^y + \partial_y m_{\sf T_1}^x 
	\end{pmatrix} \bigg|^2
	+ g_1 \bigg|\begin{pmatrix}
	\partial_x m_{\sf E}^1 \\
	-\frac{1}{2} \partial_y m_{\sf E}^1 + \frac{\sqrt{3}}{2} \partial_y m_{\sf E}^2 \\
	-\frac{1}{2} \partial_z m_{\sf E}^1 - \frac{\sqrt{3}}{2} \partial_z m_{\sf E}^2 
	\end{pmatrix} \bigg|^2
	\nonumber \\
	&& \qquad
	+ g_2 \bigg|\begin{pmatrix}
	\partial_x m_{\sf E}^2 \\
	-\frac{1}{2} \partial_y m_{\sf E}^2 - \frac{\sqrt{3}}{2} \partial_y m_{\sf E}^1 \\
	-\frac{1}{2} \partial_z m_{\sf E}^2 + \frac{\sqrt{3}}{2} \partial_z m_{\sf E}^1 
	\end{pmatrix} \bigg|^2 
	\nonumber \\
	&& \qquad
	+ k_1 
	\begin{pmatrix}
	\partial_x m_{\sf E}^1 \\
	-\frac{1}{2} \partial_y m_{\sf E}^1 + \frac{\sqrt{3}}{2} \partial_y m_{\sf E}^2 \\
	-\frac{1}{2} \partial_z m_{\sf E}^1 - \frac{\sqrt{3}}{2} \partial_z m_{\sf E}^2 
	\end{pmatrix}
	\cdot 
	\begin{pmatrix}
	\partial_y m_{\sf T_1}^z + \partial_z m_{\sf T_1}^y \\
	\partial_z m_{\sf T_1}^x + \partial_x m_{\sf T_1}^z \\
	\partial_x m_{\sf T_1}^y + \partial_y m_{\sf T_1}^x 
	\end{pmatrix} 
	\bigg]
	\label{eq:F0-GL}
	\end{eqnarray}
	$F_0$ includes all bilinear terms in ${\bf m}_{\sf E}$, ${\bf m}_{\sf T_1}$ and their first-order spatial derivatives allowed by the time-reversal,
	inversion and the point group symmetry of the lattice.

	After Fourier transformation Eq. \ref{eq:F0-GL} can be represented as
	\begin{eqnarray}
	&&F_0 = \frac{1}{2} \int d^3 {\bf q} \  M_{\alpha}(-{\bf q}) V_{\alpha \beta}({\bf q}) M_{\beta}({\bf q})
	\end{eqnarray}
	where ${\bf M}$ is a five-component vector into which we collect both $ {\bf m}_{\sf T_1},  {\bf m}_{\sf E}$
	and $V_{\alpha \beta}({\bf q})$ is a $5\times5$ coupling matrix.

	To describe the close competition between FM and 
	AFM phases we assume $c_{\sf E}=c_{\sf T_1}=c$.
	The parameters $f_i$, $g_i$, $k_i$ are chosen such that the eigenspectrum of $V({\bf q})$ (\ref{eq:F0-GL})
	will feature low lying flat modes along the $\{111\}$ directions. This reflects the microscopic
	physics of the FM/AFM phase boundary where such a 
	mode is known to emerge \cite{Yan2017}. This is built
	into the theory by setting $k_1=\pm\sqrt{12 (f_2 + \frac{1}{3} f_3) (g_1 + g_2)}$.
	The low-lying mode thus obtained is two-fold degnerate by symmetry.

	To describe the dynamics, we use a Langevin equation,
	of a standard form appropriate for systems in which
	the order parameters are not conserved quantities
	\cite{Hohenberg77}:
	\begin{eqnarray}
	\hbar \partial_t M_{\alpha} ({\bf r}, t) = - \Gamma \frac{\delta F_0}{\delta M_{\alpha} ({\bf r}, t) } +
	\theta_{\alpha} ({\bf r}, t) + \Gamma h_{\alpha}({\bf r},t)
	\label{eq:Langevin}
	\end{eqnarray}
	$\Gamma$ is a dimensionless phenomenological parameter describing the dissipation in the system.
	The first term in Eq. (\ref{eq:Langevin}) favours relaxation towards a state minimizing $F_0$.
	$h_{\alpha}$ is an external field coupling to the order parameter components $M_{\alpha}$, which we include for the purpose of defining susceptibilities.
	$\theta_{\alpha}$ is a stochastic, Langevin noise field, which models the coupling of $M_{\alpha}$
	to short-wavelength modes which don't appear explicitly in the phenomenological theory.
	$\theta_{\alpha}$ vanishes on average and is uncorrelated in space and time:
	$\langle \theta_{\alpha}  ({\bf r}, t) \ \theta_{\alpha'} ({\bf r}', t') \rangle=
	2 k_B T \Gamma \delta({\bf r} - {\bf r}') \delta(t-t') \delta_{\alpha \alpha'}$
	The magnitude of the fluctuations of $\theta_{\alpha}$ is fixed by the requirement of thermal equilibrium (fluctuation-dissipation theorem).
	
	The a.c. susceptibility follows from Eq. (\ref{eq:Langevin})
	\begin{eqnarray}
	\chi_{\alpha \beta} ({\bf q}, \omega) =
	\sum_{\lambda} (U^T)_{\alpha \lambda} ({\bf q}) 
	\frac{\Gamma }{\Gamma \epsilon_{\lambda} ({\bf q}) - i \hbar\omega}
	U_{\lambda \beta} ({\bf q}) 
	\label{eq:im_chi_1}
	\end{eqnarray}
	where $U({\bf q})$ is the orthogonal matrix that diagonalizes $V({\bf q})$
	and $\epsilon_{\lambda}({\bf q})$ are the eigenvalues.
	
	We now consider momenta along $(1,1,1)$:
	$
	{\bf q}=\frac{1}{\sqrt{3}}q_{1 1 1}  (1,1,1)$
	$V({\bf q})$ has two flat modes along this direction with eigenvalue:
	\begin{eqnarray}
	\epsilon_{1,2} (q_{111}) = c
	\label{eq:c}
	\end{eqnarray}
	with $c$ being the coefficient in front of ${\bf m}_{\sf E}$ and ${\bf m}_{T_1}$ in the Ginzburg-Landau
	theory [Eq. (\ref{eq:F0-GL})].
	
	Let us then suppose these modes collapse to zero at some temperature $T_{\rm rod}$
	\begin{eqnarray}
	c = a k_B (T- T_{\rm rod}).
	\label{eq:mode_collapse}
	\end{eqnarray}
	
	Approaching $T_{\rm rod}$, these  modes will dominate the susceptibility, and the other modes can
	be neglected in the sum in Eq. (\ref{eq:im_chi_1}):
	\begin{eqnarray}
	\chi_{\alpha \beta} (q_{111}, \omega)  =
	\sum_{\lambda=1,2}
	\frac{ (U^T)_{\alpha \lambda} ({\bf q}) 
		U_{\lambda \beta} ({\bf q})
		\Gamma \hbar \omega}{\Gamma a k_B (T-T_{\rm rod}) -  i \hbar\omega}
	.
	\nonumber \\
	\label{eq:im_chi_2}
	\end{eqnarray}
	
	The real part of the susceptibility is
	then a Lorentzian:
	\begin{eqnarray}
	{\rm Re}[\chi_{\alpha \beta} (q_{111}, \omega)] =
	\sum_{\lambda=1,2}
	\frac{ (U^T)_{\alpha \lambda} ({\bf q}) 
		U_{\lambda \beta} ({\bf q})
		\Gamma^2 a k_B (T-T_{rod})}{\Gamma^2 a^2 k_B^2 (T-T_{\rm rod})^2 + \hbar^2\omega^2} 
	\nonumber \\
	\end{eqnarray}
	which allows us to read off a relaxation time:
	\begin{eqnarray}
	\tau_{\rm rod}=\frac{\hbar}{\Gamma a k_B (T-T_{\rm rod})}
	\end{eqnarray}
	
	The structure factor is related to the imaginary part of the susceptibility
	\begin{eqnarray}
	{\rm Im}[\chi_{\alpha \beta} (q_{111}, \omega)] =
	\sum_{\lambda=1,2}
	\frac{ (U^T)_{\alpha \lambda} ({\bf q}) 
		U_{\lambda \beta} ({\bf q})
		\Gamma \hbar \omega}{\Gamma^2 a^2 k_B^2 (T-T_{\rm rod})^2 + \hbar^2\omega^2}
	\end{eqnarray}
	
	The dynamical structure factor is then obtained using a fluctuation-dissipation relationship, bearing in mind that we need to use the quantum fluctuation-dissipation
	relationship, $$
	S({\bf q}, \omega)= 2  (n_{BE} (\omega) + 1 ) {\rm Im}[\chi({\bf q}, \omega)]
	$$ 
	when comparing to experiment
	and the classical relation
	$$
	S({\bf q}, \omega)= 2  \frac{k_B T}{\omega} {\rm Im}[\chi({\bf q}, \omega)].
	$$
	when comparing to simulation.

	Using the quantum fluctuation-dissipation relationship, 
	we arrive at the result from the main text:
	\begin{eqnarray}
	&&S_{\rm rod} (\omega)
	= \int_{q_1}^{q_2} dq_{111}
	S(q_{111}, \omega)
	\nonumber \\
	&&
	=
	2  (n_{BE}(\omega) + 1)\frac{1}{k_B^2 (T-T_{\rm rod})^2}
	\frac{B \hbar \omega}{R^2 + \frac{\hbar^2\omega^2}{k_B^2 (T-T_{\rm rod})^2}}
	\end{eqnarray}.
	Where 
	\begin{eqnarray}
	R=\Gamma a
	\label{eq:paramR}
	\end{eqnarray}
	and hence $\tau_{\rm rod}=\frac{\hbar}{R k_B T}$ when $T_{rod} \to 0$.
	
	\begin{figure}[t]
		\centering
		\includegraphics[width=0.45\textwidth]{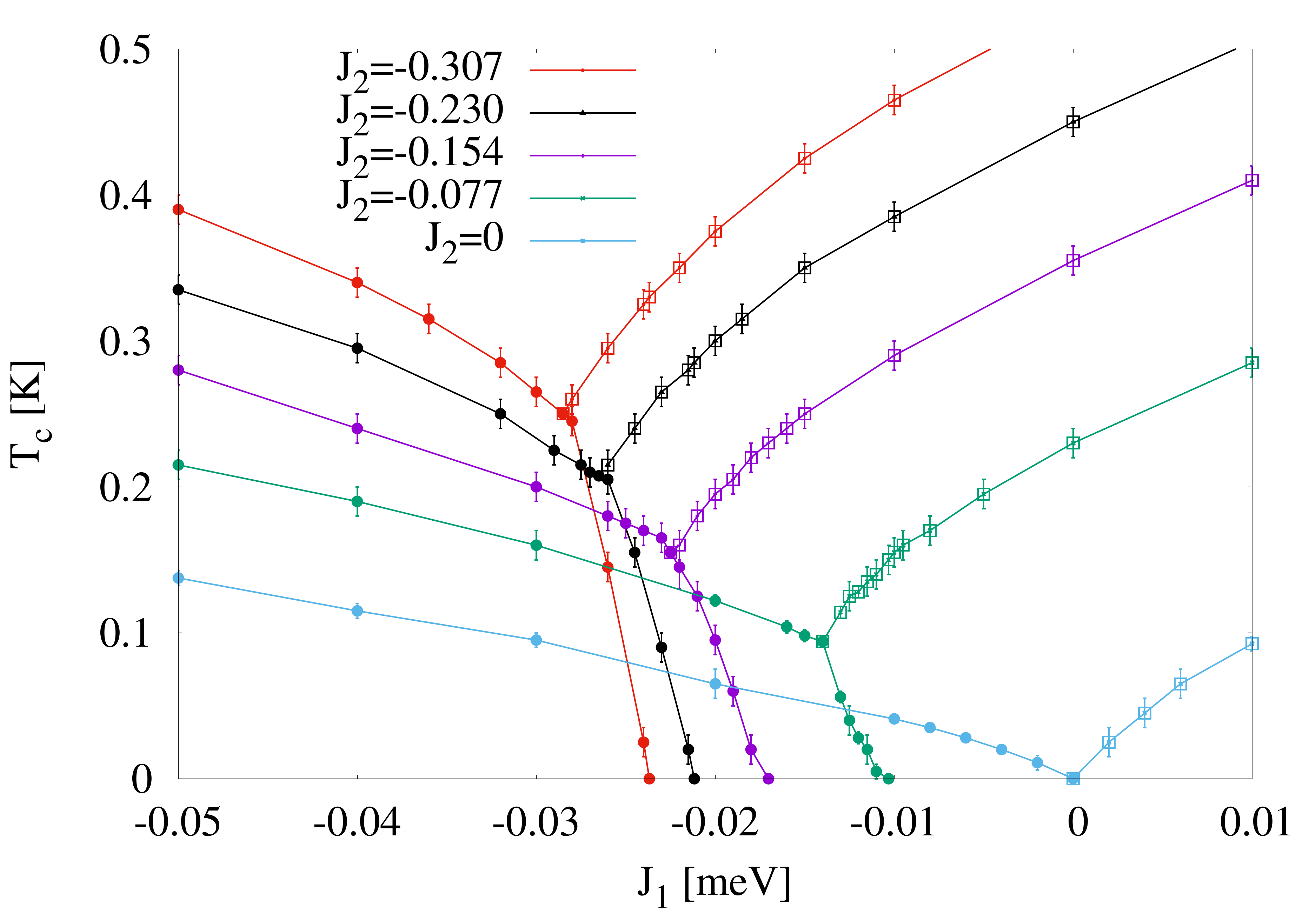}
		\caption{Finite temperature phase transitions of the anisotropic exchange model, determined from
			classical Monte Carlo simulations. This data was used to build the 3D phase diagram in main text Fig. 3. Phase transition temperatures $T_c$ are shown as a function of $J_1$, for a series of values of $J_2$, with $J_3=-0.322$meV and $J_4=-0.091 J_2$.
			Values of $J_2$ are shown in inset, in units of meV.
			Filled circles represent phase transitions into the ferromagnetic ($T_1$) phase, open squares are transitions into the antiferromagnetic ($E$) phase.
			The ordering temperature goes to
			zero approaching the spin liquid point at $J_1=J_2=J_4=0$.
		}
		\label{fig:finite_T_phase_diagram}
	\end{figure}

	\begin{figure*}
		\centering\includegraphics[width=0.9\textwidth]{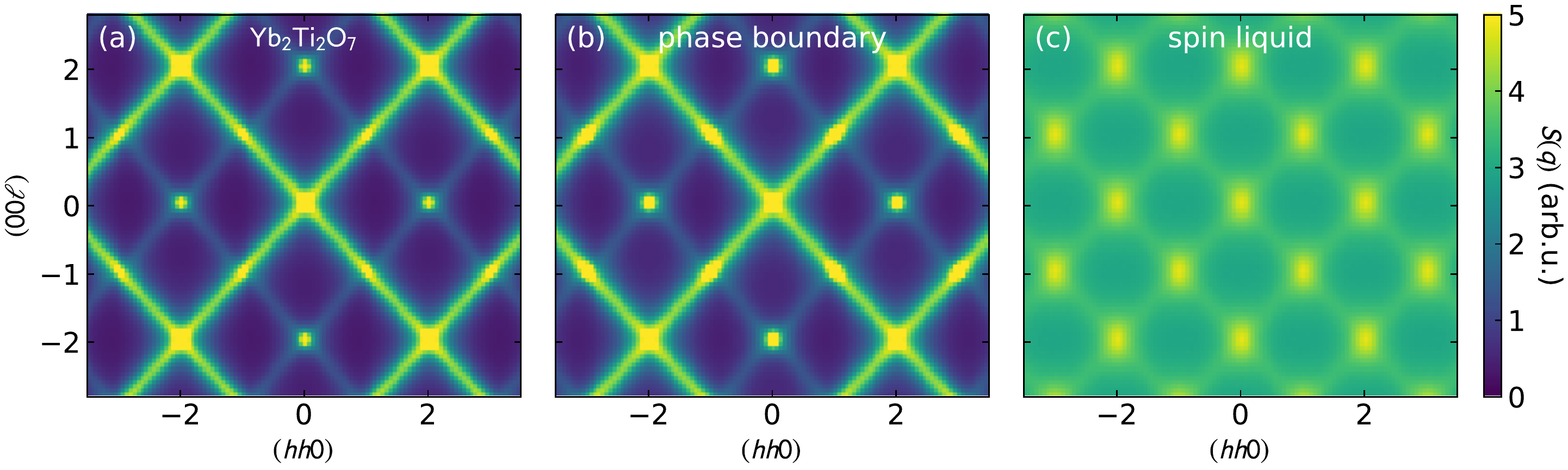}
		
		\caption{Equal time structure factor of the Hamiltonians in main text Eq. 4 calculated by Monte Carlo, using the same Hamiltonians as in main text Fig. 2: (a) of Yb$_2$Ti$_2$O$_7$, (b) on the phase boundary between FM and AFM, and (c) the pinch line spin liquid. Each panel shows the lowest temperature scattering signal above the phase transition in main text Fig 2. Note that in the the pinch line spin liquid, the weaker lines of intensity come to have intensity equal with the $(111)$ rods.
		}
		\label{flo:MDstructfact}
	\end{figure*}

	The constant of proportionality $B$
	is given by
	\begin{eqnarray}
	&&B= \Gamma
	\int_{q_1}^{q_2} dq_{111}
	\nonumber \\
	&&
	\sum_{\mu, \nu=x,y,z}
	\sum_{\lambda=1}^2
	\sum_{\alpha, \beta=1}^5
	U^T_{\alpha \lambda} (q_{111}) 
	U_{\lambda \beta} (q_{111}) 
	G_{ \mu \alpha} G^{\ast}_{\nu\beta}
	\left( 
	\delta_{\mu \nu} - \frac{q_{\mu} q_{\nu}}{q^2}
	\right)
	\nonumber \\
	\label{eq:paramB}
	\end{eqnarray}
	where $G_{ \mu \alpha}({\bf q})$ is 
	a $3\times5$ matrix that relates the 
	Fourier transform of the magnetic moment distribution, ${\bf m}({\bf q})$, to the Fourier transform of the five-component
	order parameter ${\bf M}({\bf q})$:
	\begin{eqnarray}
	m_{\mu} ({\bf q}) = \sum_{\alpha}
	G_{\mu \alpha} ({\bf q}) M_{\alpha} ({\bf q}).
	\end{eqnarray}
	
	\begin{figure*}
		\centering
		\includegraphics[width=0.9\textwidth]{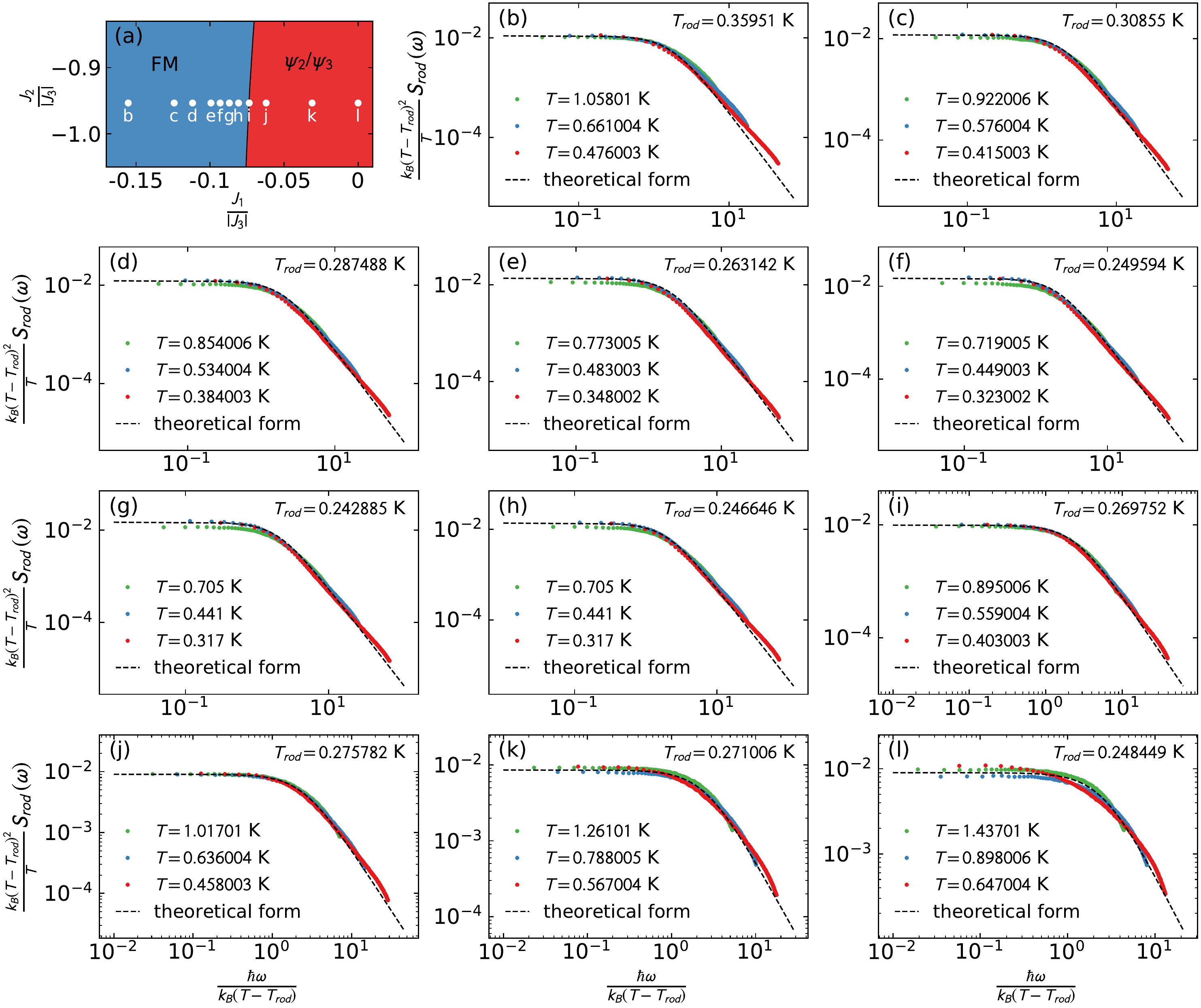}
		\caption{Collapse of $S_{\rm rod} (\omega)$ when tuning across the phase boundary between FM and AFM phases in simulation.
			(a) Section of phase diagram, with $J_3<0$ and $J_4=-0.091 J_2$ showing the location of the parameter sets used for the calculations in (b)-(l). 
			(b)-(f) Scaling collapse of MD data at three different temperatures for each parameter set. $T_{\rm rod}$ has been optimized for each set to give the best collapse.
			The parameter set in panel (i), which is tuned to lie on the FM/AFM boundary gives the most precise scaling collapse.
			The parameter set in panel  (h) corresponds with $\rm Yb_2Ti_2O_7$ \cite{scheie2020multiphase}.
			The collapse gets worse moving away from the phase boundary in both directions.
		}
		\label{fig:scaling_summary}
	\end{figure*}
	
	For comparison to simulations, combining Eq. (\ref{eq:im_chi_2}) with the classical fluctuation-dissipation relation gives:
	\begin{eqnarray}
	&&S^{\rm cl}_{\rm rod} (\omega)
	=
	\frac{k_B T}{k_B^2 (T-T_{\rm rod})^2}
	\frac{A}{W^2 + \frac{\hbar^2\omega^2}{k_B^2 (T-T_{\rm rod})^2}}
	\end{eqnarray}
	with $A$ and $W$ being non-universal parameters:
	\begin{eqnarray}
	&&A= 2 \Gamma \hbar
	\int_{q_1}^{q_2} dq_{111}
	\sum_{\mu, \nu=x,y,z}
	\sum_{\lambda=1}^2
	\sum_{\alpha, \beta=1}^5
	\nonumber \\
	&& \qquad \qquad
	U^T_{\alpha \lambda} (q_{111}) 
	U_{\lambda \beta} (q_{111})
	G_{ \mu \alpha} G^{\ast}_{\nu\beta}
	\left( 
	\delta_{\mu \nu} - \frac{q_{\mu} q_{\nu}}{q^2}
	\right)
	\label{eq:paramA}
	\\
	&&W=\Gamma a.
	\label{eq:paramW}
	\end{eqnarray}

	Comparison of Eqs. (\ref{eq:paramR}), (\ref{eq:paramB}), (\ref{eq:paramA})
	and (\ref{eq:paramW}) suggests
	$A=2\hbar B$ and $W=R$.
	We retain, however, a separate notation 
	to emphasise that we treat these as non-universal, phenomenological, 
	parameters which can be very different between
	the semi-classical and quantum systems.
	We use $A$ and $W$ to refer to the fit
	parameters for the semi-classical
	scaling and $B$ and $R$ to refer to the fit
	parameters for the experiment.
	
	\section{Monte Carlo and Molecular
		Dynamics Simulations}

	\subsection{Microscopic Model}
	
	We simulate the nearest-neighbor
	anisotropic exchange model on the pyrochlore lattice
	\begin{eqnarray}
	H_{\rm ex}=
	\sum_{\langle ij \rangle}
	\sum_{\alpha \beta} J_{ij}^{\alpha \beta} S_i^{\alpha} S_j^{\beta}.
	\label{eq:microscopic_model}
	\end{eqnarray}
	In the most general symmetry allowed 
	model \cite{Curnoe_2007, Ross_Hamiltonian, Yan2017}
	there are six distinct interaction matrices ${\bf J}_{ij}$, corresponding to each of the six bonds within a  pyrochlore tetrahedron.
	All tetrahedra in the lattice then have the same set of six interaction matrices.
	The six matrices are constrained by
	point group symmetries and depend on four independent parameters.
	Labelling the sites in a tetrahedron 
	$0,1,2,3$, the exchange matrices are:
	\begin{eqnarray}
	&{\bf J}_{01} 
	= \begin{pmatrix}
	J_2 & J_4 &J_4 \\
	-J_4 & J_1 &J_3 \\
	-J_4 & J_3 &J_1
	\end{pmatrix} 
	\quad
	&{\bf J}_{02} 
	= \begin{pmatrix}
	J_1 & -J_4 & J_3 \\
	J_4 & J_2 & J_4 \\
	J_3 & -J_4 & J_1
	\end{pmatrix}    \nonumber\\
	&{\bf J}_{03} 
	= \begin{pmatrix}
	J_1 & J_3 & -J_4 \\
	J_3 & J_1 & -J_4 \\
	J_4 & J_4 & J_2
	\end{pmatrix} 
	\quad
	&{\bf J}_{12} 
	= \begin{pmatrix}
	J_1 & -J_3 & J_4 \\
	-J_3 & J_1 & -J_4 \\
	-J_4 & J_4 & J_2
	\end{pmatrix}    \nonumber\\
	&{\bf J}_{13} 
	= \begin{pmatrix}
	J_1 & J_4 & -J_3 \\
	-J_4 & J_2 & J_4 \\
	-J_3 & -J_4 & J_1
	\end{pmatrix} 
	\quad
	&{\bf J}_{23} 
	= \begin{pmatrix}
	J_2 & -J_4 & J_4 \\
	J_4 & J_1 & -J_3 \\
	-J_4 & -J_3 & J_1
	\end{pmatrix} \nonumber\\ 
	\label{eq:Jij}
	\end{eqnarray}

	\subsection{Monte Carlo (MC)}

	Monte Carlo simulations are performed on systems of classical Heisenberg spins with $N=16L^{3}$ sites, where $L^{3}$ is the number of cubic unit cells. The spin length is $|S|=1/2$. Several update algorithms are used together: the heatbath method, over-relaxation and parallel tempering. Parallel tempering is done every 100 Monte Carlo steps (MCS) and overrelaxation is done at every MCS. Thermalization is made in two steps: first a slow annealing from high temperature to the temperature of measurement $T$ during $t_{e}$ MCS followed by $t_{e}$ MCS at temperature $T$. After thermalization, measurements are done every 10 MCS during $t_{m}=10~t_{e}$ MCS.
	
	The characteristics of our simulations are typically:
	\begin{itemize}
		\item $L = 8$,
		\item $10^6 \leq t_{m} \leq 10^7$ MCS,
		\item 100 different temperatures (regularly spaced) for parallel tempering.
	\end{itemize}

	\subsubsection{Finite temperature phase diagram}
	
	We have used classical Monte Carlo to determine the finite temperature phase diagram of the nearest neighbor exchange model on a path through parameter space interpolating between the parameters determined for $\rm Yb_2Ti_2O_7$ in \cite{scheie2020multiphase} and the pinch-line spin liquid \cite{benton16-pinchline}.
	
	The results are shown on a three-dimensional phase diagram in Fig. 3 of the main text, and in a two-dimensional projection of parameter space in Fig. \ref{fig:finite_T_phase_diagram} and show a continuous suppression of the transition temperature as the spin liquid is approached.
	
	The equal-time structure factors of the lowest temperatures of the three parameter sets in main text Fig. 2 have also been calculated from classical Monte Carlo and are plotted in Fig. \ref{flo:MDstructfact}.

	\subsubsection{Generation of configurations for Molecular Dynamics (MD) simulations}

	The initial configurations for the MD simulations of spin dynamics are also
	generated using a Monte Carlo method, using a different algorithm than the
	simulations described above.
	
	The initial configurations used for the dynamics are generated from a Monte
	Carlo sampling based on the heat bath update. To speed up the update process, we
	use the property that all spins of a given sublattice of the pyrochlore lattice
	can be updated in parallel because they do not interact with each other through
	first neighbors interactions.
	
	We choose a unit cell of 16 spins to carry out the simulations. In this
	configuration, we find that (i) 4 sublattices can be updated in a single step at
	any given time, (ii) there are 24 different combinations of four independent
	sublattices. A full lattice update corresponds to a selection of 4 sets of
	sublattices taken randomly. All spins of each sublattice set is updated using
	the local heat bath algorithm. This update, which we call a lattice sweep, is the
	basic update of these simulations.
	
	We use simulated annealing to go from the high temperature paramagnetic
	regime to the target temperature of the spin configurations. The other
	parameters for the simulated annealing are the following:
	
	\begin{itemize}
		\item 16 different replicas of the same spatial size all start from the high
		temperature paramagnetic regime of temperature $T_{\rm start}$.
		\item Each replica is slowly cooled down to its respective measurement
		temperature $T^i$; $i$ is the replica index; with a temperature step given by
		$\delta T = (T_{\rm start} - T^i) / N$, $N = 50000$ being the number of
		intermediate temperatures.
		\item 50 lattice sweeps are applied to all replica between temperature updates.
	\end{itemize}
	
	Parallel tempering is applied before the thermalization cycle. We apply 500
	parallel tempering steps and 2500 lattices sweeps to each configurations between
	each parallel tempering swap.
	
	Finally the full set of configurations is thermalized at their respective target
	temperature with 125000 lattice sweeps. 250 spin configurations are regularly
	extracted during the calculation of the specific heat and order parameters.
	
	The two Monte Carlo codes used in this work are the foundation of all numerical
	results reported in Ref.~\cite{taillefumier17} and are constantly compared using
	thermodynamics.

	\subsection{Molecular Dynamics (MD)}
	
	The Heisenberg equation of motion derived from the microscopic model, Eq.~(\ref{eq:microscopic_model}), was numerically integrated using an explicit,  $8^{th}$--order Runge--Kutta scheme, with Dormand and Prince coefficients given in \cite{Hairer1992}.
	These results  were validated by comparison with numerical integration carried out using the more demanding implicit Runge Kutta method, with no major differences found over time scales of interest.
	
	250 spin configurations of $N = 16^4$ spins are sampled from the thermalized
	ensemble given by Monte Carlo simulations for the target temperatures. The
	interval of integration is fixed to $2048 \> J_3^{-1}$ to cover the full
	spectrum while the time step is set to $\delta t = 1/10 \> J_3^{-1}$. With these
	parameters the total energy per spin drifts by an amount less than $10^{-8} J_3^{-1}$ over the
	entire interval of integration.
	
	The effect of the sharp boundaries of the time window can be mitigated by
	multiplying the signal with a Dolph-Chebyshev window before calculating the
	time-dependent Fourier transform. The dynamical structure factor is then
	evaluated and averaged over the different spin
	configurations.

	\subsubsection{Behavior of dynamical scaling collapse tuning across phase boundary}
	
	Fig. \ref{fig:scaling_summary} shows the scaling collapse of the data for $S_{\rm rod}(\omega)$ for a series of parameter sets crossing the FM/AFM phase boundary.
	The collapse is best in Fig. \ref{fig:scaling_summary}(i), which is tuned to the phase boundary and gets worse when tuning away.

\end{document}